\documentclass[conference]{IEEEtran}

\usepackage[T1]{fontenc}
\usepackage[utf8]{inputenc}
\usepackage{amsmath,amssymb}
\usepackage[capitalize,noabbrev]{cleveref} %
\usepackage{microtype}
\usepackage[svgnames,x11names]{xcolor}
\usepackage{graphicx}
\usepackage{tabularx}
\usepackage{multirow}
\usepackage[shortlabels,inline]{enumitem}
\usepackage{xspace}
\usepackage{balance}
\usepackage[colorinlistoftodos,prependcaption,textsize=tiny]{todonotes} %
\usepackage{booktabs}
\usepackage{comment}

\usepackage[abbreviate=true, dateabbrev=true, alldates=year, isbn=false, doi=false, urldate=comp, url=true, maxbibnames=1, backref=false, backend=biber, style=numeric-comp, language=american, giveninits=true, sortcites=true, sorting=none]{biblatex}
\usepackage{citation-cleaner}

\AtEveryBibitem{\clearlist{location}} %
\AtEveryBibitem{\clearfield{series}} %
\AtEveryBibitem{\clearlist{language}} %

\RemoveBibField{weyns2022:preliminary}{number}
\RemoveBibField{weyns2022:preliminary}{institution}
\RemoveBibField{weyns2022:activforms}{number}
\RemoveBibField{weyns2022:activforms}{institution}
\RemoveBibField{duarte2022:evaluation}{eprint}
\RemoveBibField{duarte2022:evaluation}{institution}

\RemoveBibField{satoh2012:bioinspired}{editor}
\RemoveBibField{tamura2013:practical}{editor}
\RemoveBibField{delemos2017:software}{editor}
\RemoveBibField{raibulet2017:overview}{editor}
\RemoveBibField{tomforde2018:comparing}{editor}
\RemoveBibField{filieri2013:probabilistic}{editor}
\RemoveBibField{eberhardinger2018:measuring}{editor}
\RemoveBibField{dias2020:visual}{editor}
\RemoveBibField{gallet2010:model}{editor}

\RemoveBibField{reinecke2010:evaluating}{pages}
\RemoveBibField{dasilva2021:catalog}{pages}
\RemoveBibField{hussein2013:scenariobased}{pages}
\RemoveBibField{camara2012:evaluation}{pages}
\RemoveBibField{gerostathopoulos2021:how}{pages}
\RemoveBibField{oquendo2017:software}{pages}
\RemoveBibField{ma2019:testing}{pages}
\RemoveBibField{passini2022:design}{pages}
\RemoveBibField{gadelha2020:scen}{pages}
\RemoveBibField{iglesia2015:mape}{pages}
\RemoveBibField{almeida2011:benchmarking}{pages}
\RemoveBibField{almeida2011:benchmarking}{publisher}
\RemoveBibField{scheerer2021:reliability}{pages}
\RemoveBibField{scheerer2021:reliability}{publisher}
\RemoveBibField{filieri2013:probabilistic}{pages}
\RemoveBibField{filieri2013:probabilistic}{publisher}
\RemoveBibField{desousa2019:quality}{pages}
\RemoveBibField{desousa2019:quality}{publisher}

\RemoveBibField{barr2001:quagmire}{publisher}
\RemoveBibField{barr2001:quagmire}{pages}

\RemoveBibField{bass2013:software}{edition}

\newcommand{\LM}[2][]{\todo[linecolor=orange,backgroundcolor=orange!40,bordercolor=orange,#1]{#2}} %
{\newsavebox{\myStore}\LM{#1}\begin{lrbox}{\myStore}\hspace*{-1ex}\begin{minipage}{\textwidth-1ex}}%
		{\end{minipage}\end{lrbox}\LM[inline,nolist]{\usebox{\myStore}}}

\newcommand{\MN}[2][]{\todo[linecolor=green,backgroundcolor=green!25,bordercolor=green,#1]{#2}}

{\newsavebox{\myStore}\MN{#1}\begin{lrbox}{\myStore}\hspace*{-1ex}\begin{minipage}{\textwidth-1ex}}%
		{\end{minipage}\end{lrbox}\MN[inline,nolist]{\usebox{\myStore}}}

\newcommand{\SM}[2][]{\todo[linecolor=blue,backgroundcolor=blue!25,bordercolor=blue,#1]{#2}}

{\newsavebox{\myStore}\SM{#1}\begin{lrbox}{\myStore}\hspace*{-1ex}\begin{minipage}{\textwidth-1ex}}%
		{\end{minipage}\end{lrbox}\SM[inline,no list]{\usebox{\myStore}}}

\newcommand{\MA}[2][]{\todo[linecolor=Purple1,backgroundcolor=Purple1!25,bordercolor=Purple1,#1]{#2}}

{\newsavebox{\myStore}\MA{#1}\begin{lrbox}{\myStore}\hspace*{-1ex}\begin{minipage}{\textwidth-1ex}}%
		{\end{minipage}\end{lrbox}\MA[inline,no list]{\usebox{\myStore}}}

\newlist{researchquestions}{enumerate}{2}
\setlist[researchquestions,1]{label=\textit{\textbf{RQ\arabic*:}},ref=RQ\arabic*,left=\parindent}
\setlist[researchquestions,2]{label=\textit{\alph*)},ref=\theresearchquestionsi(\alph*),leftmargin=0pt}

\newcommand{\head}[1]{\par\noindent\textbf{#1:}\space}

\title{On Evaluating Self-Adaptive and Self-Healing Systems using Chaos Engineering}

\author{%
	\IEEEauthorblockN{%
		Moeen Ali Naqvi \quad\quad%
		Sehrish Malik \quad\quad%
		Merve Astekin \quad\quad%
		Leon Moonen\\[0.8ex]}
	\IEEEauthorblockA{Simula Research Laboratory, Oslo, Norway}
	\IEEEauthorblockA{Email:
		\{moeen,sehrish,merve\}@simula.no, 
		leon.moonen@computer.org}%
}

\makeatletter
\def\ps@IEEEtitlepagestyle{%
  \def\@oddfoot{\mycopyrightnotice}%
  \def\@evenfoot{}%
}
\def\mycopyrightnotice{%
  \hspace*{3mm}\includegraphics[width=2cm]{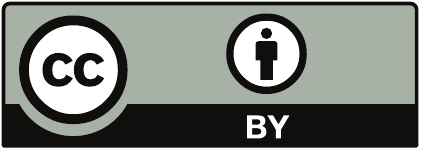}%
  \hspace*{2mm}\raisebox{2.5mm}{%
   	  \parbox{\columnwidth}{\footnotesize This work is licensed under a Creative Commons \\ Attribution 4.0 International (CC BY 4.0) license.}%
   	  \hspace*{-68pt}\mbox{1}\hspace{20pt}\fbox{\parbox{.9\columnwidth}{\footnotesize\textsl{Accepted for publication in the 3rd IEEE International Conference on Autonomic Computing and Self-Organizing Systems (ACSOS 2022).}}}%
  }%
  \gdef\mycopyrightnotice{}%
}
\makeatother

\begin{document}	

\maketitle

\pagestyle{plain}

\noindent%

\begin{abstract}
With the growing adoption of self-adaptive systems in various domains,
there is an increasing need for strategies to assess their correct behavior.
In particular self-healing systems, which aim to provide resilience and fault-tolerance,
often deal with unanticipated failures in critical and highly dynamic environments.   
Their reactive and complex behavior makes it challenging to assess
if these systems execute according to the desired goals.
Recently, several studies have expressed concern about the lack of systematic evaluation methods for self-healing behavior.

In this paper, we propose CHESS, 
an approach for the systematic evaluation of self-adaptive and self-healing systems that builds on \emph{chaos engineering}.
Chaos engineering is a methodology for subjecting a system to unexpected conditions and scenarios.
It has shown great promise in helping developers build resilient microservice architectures and cyber-physical systems. 
CHESS turns this idea around by using chaos engineering to evaluate \emph{how well} a self-healing system can withstand such perturbations.
We investigate the viability of this approach through an exploratory study on a self-healing smart office environment.
The study helps us explore the promises and limitations of the approach, as well as identify directions where additional work is needed.
We conclude with a summary of lessons learned.

\end{abstract}

 \begin{IEEEkeywords}
self-healing,
resilience, 
chaos engineering, 
evaluation, 
exploratory study
 \end{IEEEkeywords}

\section{Introduction}

\noindent
There is a growing interest in the research of self-adaptive and self-healing systems in domains such as the internet of things (IoT), 
Infrastructure as a Service (IaaS), cyber-physical systems (CPS), and Industry 4.0, 
with some of these systems likely to be adopted into mainstream solutions~\cite{weyns2022:preliminary}. 
Systems in these domains often have to deal with uncertainty and unanticipated behavior
due to the highly dynamic environments in which they operate, 
which increases the need for providing fault tolerance and resilient behavior~\cite{ding2021:secure}.
Known strategies for achieving these qualities include monitoring, reconfiguring, redundancy, maintenance, and automated repair~\cite{bass2013:software}.
However, given the complex and dynamic nature of the domains in which these systems operate, it is challenging to anticipate all possible scenarios.
Therefore, there is a growing trend towards systems that are capable of making dynamic decisions at runtime.

Several studies have expressed concern about the lack of systematic evaluation of self-adaptive systems (SAS) and self-healing systems (SHS)~\cite{gerostathopoulos2021:how,passini2022:design,desousa2019:quality}. 
A systematic mapping study of self-adaptive service-oriented applications shows 
that only 7 out of 60 studies deal with the evaluation of previously developed applications~\cite{passini2022:design}.
In addition, there is a lack of automated tools to support evaluations based on runtime measures, 
and most studies concern evaluations focusing on the models used to design the system~\cite{desousa2019:quality}.
These models estimate the type of failure and respective repair actions at either design or deployment time, 
which usually misses some crucial scenarios that a system can face under operation. 
Runtime models overcome some of these limitations by leveraging first-class abstractions of the runtime system, 
but they come with their own challenges, such as the need for model creation and maintenance~\cite{ghahremani2020:evaluation}.

At a high-level, self-adaptive and self-healing systems can be seen as comprising a \emph{managed system} that is controlled by a \emph{managing system}.  
Although there is a considerable body of work on \emph{evaluating managed systems}, far less attention is given to \emph{evaluating managing systems}, the topic of this paper.
Several aspects of self-adaptive systems make it challenging to systematically evaluate their behavior.
First, a wide variety of approaches can be used to engineer the managing systems, 
ranging from static, reactive, parametric solutions to dynamic, proactive, structural solutions~\cite{wong2022:selfadaptive}. 
Second, parts of the managing system can be realized using black-box components, e.g., from control engineering~\cite{iglesia2015:mapek}, 
bio-inspired solutions~\cite{satoh2012:bioinspired}, or reinforcement learning~\cite{cioara2010:reinforcement}.
Moreover, the adaptation strategy plays an essential role, and there are many levels where adaptation can occur, including the system software, 
specific components in the system, communication between components, or the context itself~\cite{krupitzer2016:comparison}.
Finally, when SAS/SHS take dynamic decisions at runtime, they can exhibit emergent behaviors not seen or conceived before~\cite{oquendo2017:software}.

\head{Contributions} 
The key contributions of this work include:
\begin{itemize}
\item We survey the state-of-the-art in evaluating self-adaptive and self-healing systems, 
	highlighting the main quality attributes and distinguishing the main evaluation approaches. 
\item We propose \emph{CHESS}, an approach for the systematic evaluation of self-adaptive and self-healing systems that builds on chaos engineering principles. 
	The approach systematically perturbates the system-under-evaluation and records how the system responds to those perturbations.
\item We present the experimental design for evaluating distributed SHS based on microservices and discuss common failure scenarios and their mapping to quality attributes. 
\item We examine the viability of CHESS through an exploratory study that evaluates a self-healing smart office application.
\item We discuss the \emph{lessons learned} while conducting the exploratory study, 
which includes challenges w.r.t. observability, chaos experiments at the functional level, and limiting the cascading effects of chaos experiments. 
\end{itemize}

\section{Background}

\noindent
When studying the literature on self-adaptive and self-healing systems, some confusion can arise around the 
terms \emph{evaluation}, \emph{testing}, \emph{assurance}, \emph{verification}, and \emph{validation}, 
as these are used with both different and overlapping meanings in various relevant studies.
To ensure a common understanding, we review some of the relevant notions from the literature, 
and establish a working definition of what we mean with the term \emph{evaluation}.

Tamura \emph{et al.}~\cite{tamura2013:practical} discuss the notion of runtime verification and validation of 
self-adaptive software systems in comparison to V\&V in software engineering.
A concept of \emph{viability zone} is introduced as a set of possible states in which the system is not compromised.
The goal of V\&V is to keep the system inside its viability zone.
Verification and validation tasks are added as common elements with each component of a MAPE-K feedback loop.
The notion of \emph{assurance} is quite often used for related concepts in the domain of 
self-adaptive systems~\cite{delemos2017:software}.
Provisioning assurances aim to provide specific guarantees about the functionality 
and quality of the self-adaptive system and to manage uncertainties.
Several techniques to provide guarantees under uncertainties are discussed by Weyns~\cite{weyns2021:introduction}. 
However, providing evidence for the value of self-adaptive systems is still considered one of the biggest challenges. 

For defining evaluation, we build on Barr's work that examines testing and evaluation in the context of V\&V for new domains~\cite{barr2001:quagmire}.
The study distinguishes testing and evaluation based on several factors:
Testing generally takes place earlier in the development lifecycle.
It focuses on identifying and correcting faults in the implemented code 
and involves code coverage determined based on implementation details.
In contrast, evaluation usually occurs later in the development lifecycle, most often after a system is complete.
It determines how well the system works and how it will perform when put into operation.
Systems are evaluated regardless of the implementation details, and the overall focus is on domain coverage. 
Therefore, we define evaluation of SAS and SHS as \emph{"an approach to determine if a system meets objectives under operation,
	identify areas in which the system performs as well as desired or predicted,
	and provide evidence to the value and applicability of the system"}.

\section{Related Work}\label{sec:relwork}

\noindent 
We survey the state-of-the-art on evaluating self-adaptive and self-healing systems and categorize it into six main topics:
\head{Reviews of evaluations in existing literature}
Gerostathopoulos \emph{et al.} investigate how studies published over the last decade at the International Symposium on Software Engineering for Adaptive and Self-Managing Systems (SEAMS) were evaluated~\cite{gerostathopoulos2021:how}.
The authors provide an in-depth analysis and characterization of how the experimental evaluations have been designed, conducted, analyzed, and packaged. 
Raibulet \emph{et al.} propose a taxonomy for structuring evaluations on self-* systems~\cite{raibulet2017:overview}. 
The taxonomy comprises elements such as the scope of the system (managed or managing), 
whether the evaluation concerns the entire software or a part of it, design time or runtime executions, adaptation types, etc.
Ghahremani \emph{et al.} present the state-of-the-art in 
evaluating the performances of self-healing systems and classifying different input types (failure models) for these systems~\cite{ghahremani2020:evaluation}.
One of their main findings is that inputs used for evaluation are often not sophisticated enough to represent real-life scenarios.
They present experiments on a simulator of mRUBiS (an exemplar of a self-adaptive marketplace that hosts an arbitrary number of shops~\cite{vogel2018:mrubis}), which show how such weak inputs can lead to incorrect conclusions from an evaluation. 

\head{Evaluation frameworks, criteria, and metrics}
Several studies have developed frameworks to guide evaluation.
The \emph{Performability framework}~\cite{meyer1980:evaluating} has been quite popular in dealing with the analysis of the fault-tolerant system. 
Villegas \emph{et al.}~\cite{villegas2011:framework} propose a framework for evaluating quality-driven SAS 
and provide a detailed mapping between adaptation properties (derived from control theory properties) and software quality attributes.
For instance, the adaptation property \emph{Robustness} is linked with the quality attributes dependability (i.e., reliability and availability) and safety. 
The \emph{adaptivity metrics framework}~\cite{reinecke2010:evaluating} is proposed for measuring the adaptivity of a computing system. 
A metric and a framework are also presented for the performance evaluation of the self-organizing mechanism~\cite{eberhardinger2018:measuring}. 
Although not directly guiding the evaluation, the SEAMS community collected a set of exemplars and reusable artifacts to facilitate reproducible research.\footnote{~http://self-adaptive.org/exemplars}

Other studies have focused on the criteria and metrics to perform the evaluation.
Self-healing benchmark~\cite{brown2005:measuring} is one of the earliest attempts to provide a mechanism to evaluate the recently introduced SHS at that time. 
They propose \emph{effectiveness score} and \emph{autonomic maturity} as the metrics to measure how effectively a system heals to disturbances, 
and how autonomic the healing response is, respectively.
Kaddoum \emph{et al.}~\cite{kaddoum2010:criteria} outline criteria for different categories for evaluating SAS and SHS including runtime evaluation. 
They propose the notion of \emph{homeostasis} and \emph{robustness ability}, defined as 
the capacity of regaining an ideal state in which the system is operating in a maximum efficient way after being perturbed,
and the system's capacity to maintain its behavior when perturbations occur, respectively. 
Almeida \emph{et al.}~\cite{almeida2011:benchmarking} propose resilience benchmarking of SAS 
through the extension of the previous works on performance and dependability benchmarks.
A quantification method for robustness in self-adaptive and self-organizing systems (SASO) is also discussed~\cite{tomforde2018:comparing}.
Recently, a catalog of 18 performance measures was extracted from 32 previous studies that evaluated SAS~\cite{dasilva2021:catalog}. 

\head{Model-based Evaluation}
Lotus@Runtime~\cite{barbosa2017:lotus} utilizes models for runtime monitoring and verification of SAS.
The tool updates the system model, which is created at design time, with the new probabilities of occurrences of each system action at runtime,
and performs runtime checks against the updated probabilistic state-based model of the system.
Hussein \emph{et al.} introduce a scenario-based approach for validating the requirements of context-aware adaptive services 
along with a technique to enumerate and generate the services' variants from their scenarios 
which are then transformed into formal models to validate against the relevant service properties~\cite{hussein2013:scenariobased}.
The tool extends the UML sequence diagram to specify service properties that must be maintained when the service adapts at runtime.
Torjusen \emph{et al.}~\cite{torjusen2014:runtime} integrate runtime verification enablers, which are models@runtime, requirements@runtime, 
dynamic context monitoring, and runtime verification component,
into the feedback loop of the ASSET project, which is an adaptive security framework for IoT in eHealth. 

\head{Metrics-based Evaluation}
Cheng \emph{et al.}~\cite{cheng2009:evaluating} evaluate a self-adaptive system implemented using the Rainbow framework~\cite{garlan2004:rainbow}.
Although the self-adaptation framework is model-based, the evaluation is based on performance metrics.
TESS is an automated performance evaluation testbed for self-adaptive and self-healing systems~\cite{porter2017:design}. 
The system collects metrics from the logs generated during execution.   
Aktas \emph{et al.}~\cite{aktas2017:provenance} propose a runtime verification mechanism 
which applies a rule-based pattern detection on the provenance metadata of the execution traces of a self-healing IoT application 
to identify faulty behaviors at runtime.
Duarte \emph{et al.}~\cite{duarte2022:evaluation} evaluate a self-healing IoT system based on Node-RED~\cite{dias2020:visual}.
They present several experiments simulating two IoT failure scenarios regarding sensor reading and timing issues. 
The experiments make use of fault injection in an instrumented version of the MQTT message broker.  
Our study uses a chaos engine instead of an instrumented MQTT broker  
and experiments with four failure scenarios where two of them overlap with their scenarios. 

\head{Model Checking}
Camara \emph{et al.}~\cite{camara2012:evaluation} propose \emph{probabilistic model checking} for evaluating the resilience of SAS. 
They collect experimental data by stimulating the system's environment and generate a model based on the aggregated execution traces. 
System properties are verified by checking the generated model against a specification.
Filieri \emph{et al.}~\cite{filieri2013:probabilistic} propose runtime probabilistic model checking for SAS.
They compare the existing approaches that used model checking and focus on the reliability and performance properties of the system. 
\emph{Runtime Quantitative Verification} (RQV) is a well-known technique 
that implements closed-loop control of SAS based on stochastic models of the system~\cite{calinescu2011:dynamic}. %
Scen@rist~\cite{gadelha2020:scen} is a scenario-based approach that extends Lotus@Runtime~\cite{barbosa2017:lotus}.
The tool collects scenario-based execution traces of an instrumented version of a SAS at runtime, 
transforms them into probabilistic state-based models, and uses model checking to check conformance against properties specified by the user. 
The ActivFORMS approach is introduced to automatically analyze the compliance with the adaptation goals of a SAS at runtime 
by utilizing automata models and statistical model checking~\cite{weyns2022:activforms}.

\head{Testing self-adaptive systems}
Two recent systematic literature reviews focus on testing self-adaptive systems: 
Siqueira \emph{et al.}~\cite{siqueira2021:testing} analyze and characterize different approaches for testing SAS, and 
Lahami \emph{et al.}~\cite{lahami2021:survey} present advances and approaches for runtime testing of dynamically adaptable and distributed systems.
King \emph{et al.}~\cite{king2011:comparative} introduce implicit self-testing of SAS. 
They propose two strategies for validating the managed systems at runtime: RV (replication with validation) and SAV (safe adaptation with validation).
RV tests adaptive changes using copies of the managed resources, whereas SAV deals directly with the managed resources.   
Other studies have used model-based testing to validate self-healing CPS.
One such study proposes a modeling framework to specify executable test models 
and an accompanying test model executor to execute these models~\cite{ma2019:modeling}.
Another proposes a \emph{fragility-oriented testing} approach that learns from test executions 
and introduces uncertainties to test the self-healing behaviors of a CPS~\cite{ma2019:testing}. 

\section{CHESS: Chaos Engineering for Evaluating Self-Adaptive and Self-Healing Systems}

\begin{figure}[b]
	\vspace*{-1ex}
	\centering
	\includegraphics[width=0.95\columnwidth]{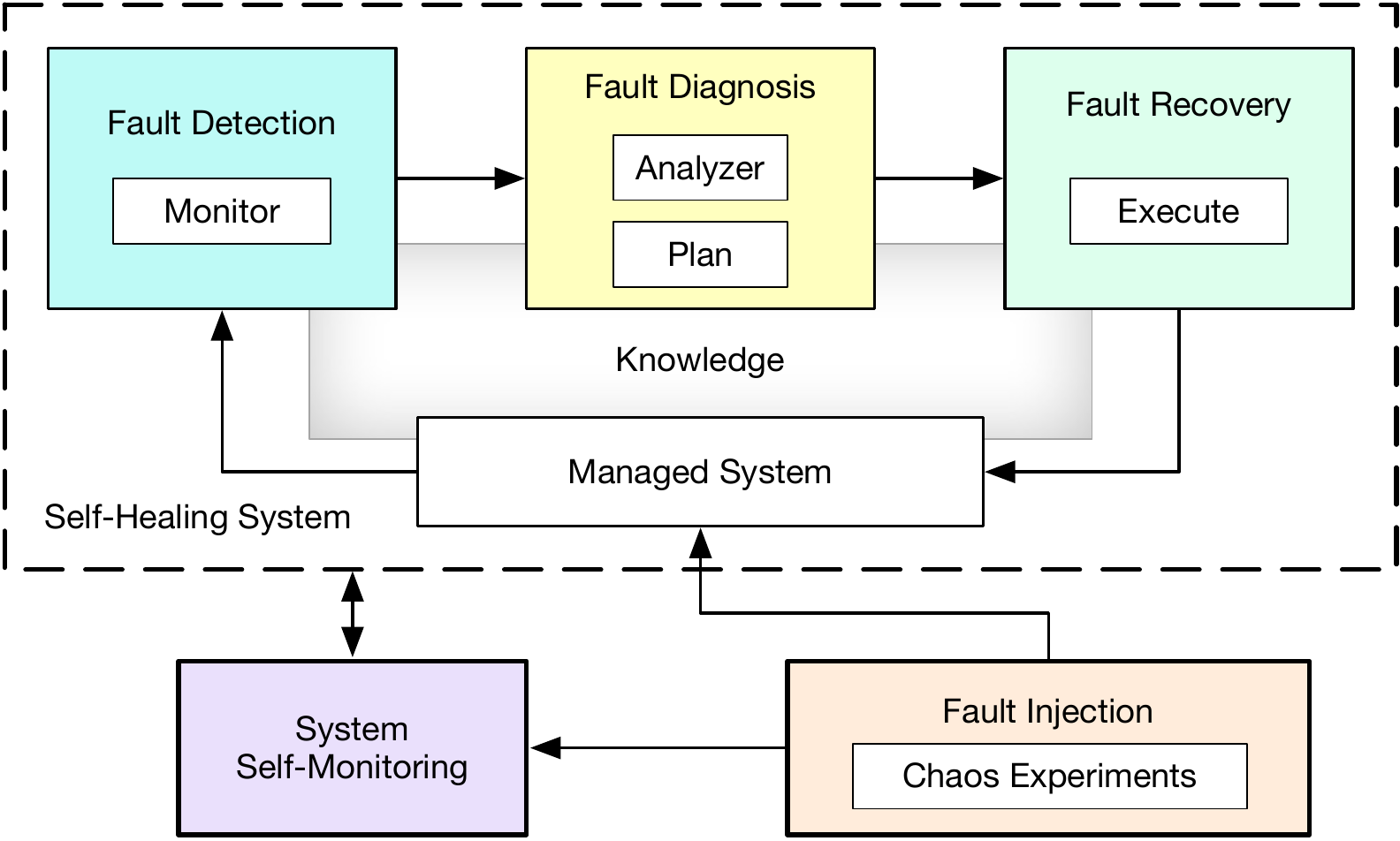}
	\caption{\label{fig:SystemOverview}Overall architecture of CHESS}
\end{figure}

\noindent 
Our survey of state-of-the-art (Section~\ref{sec:relwork}) shows that there is a shortage of generic techniques 
for the systematic evaluation of SAS and SHS based on their execution under real-life failure scenarios (as opposed to comparisons to conceptual models).
We address this gap by introducing CHESS, which evaluates SAS and SHS
through a mechanism that systematically exposes the system to faults and checks whether it can recover from these perturbations. 
The overall architecture of CHESS is shown in Figure~\ref{fig:SystemOverview}.
Faults are injected into the \emph{managed system} following the principles of chaos engineering (CE).
The self-healing system comprises a feedback loop that follows the MAPE-K reference model~\cite{kephart2003:vision}, 
reflected in the fault detection, fault diagnosis, and fault recovery and knowledge modules. 
System self-monitoring provides extensive monitoring and data collection to capture the status of the system-under-evaluation before, during, and after fault injection for further analysis. 
 
Chaos engineering is a discipline of experimenting with software systems in production-like environments 
to build confidence in their capability to withstand turbulent and unexpected conditions~\cite{basiri2016:platform}.
Improving system resilience through fault injection has been practiced for decades, 
and in recent years chaos engineering has emerged as a popular technique for conducting systematic fault injection experiments~\cite{brousse2018:use}.

\subsection{Process Flow of CHESS}

\begin{figure}[t]
	\vspace*{1ex}
	\centering
	\includegraphics[width=0.9\columnwidth]{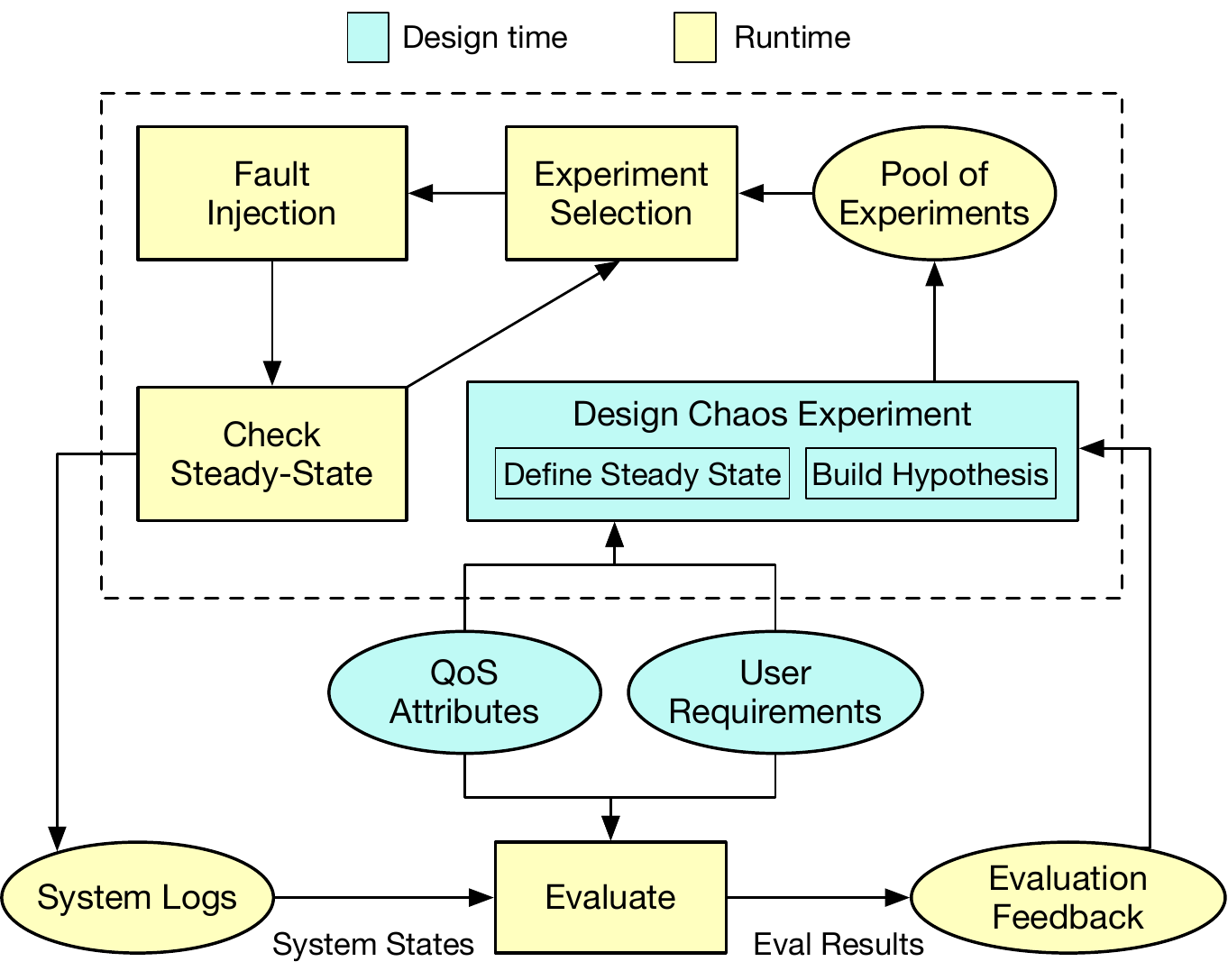}
	\caption{\label{fig:chess}High level process flow in CHESS}
	\vspace*{-2ex}
\end{figure}

\noindent
Our proposed approach leverages CE by using the chaos cycle as part of the process flow and by following the main principles of CE to design and run the chaos experiments (Figure~\ref{fig:chess}). 
The four main principles of CE are as follows: building a hypothesis around steady-state behavior, 
varying real-world events, running experiments in production and automating experiments to run continuously~\cite{basiri2016:platform}. 

\begin{figure}[t]
	\vspace*{1ex}
	\centering
	\includegraphics[width=0.95\columnwidth]{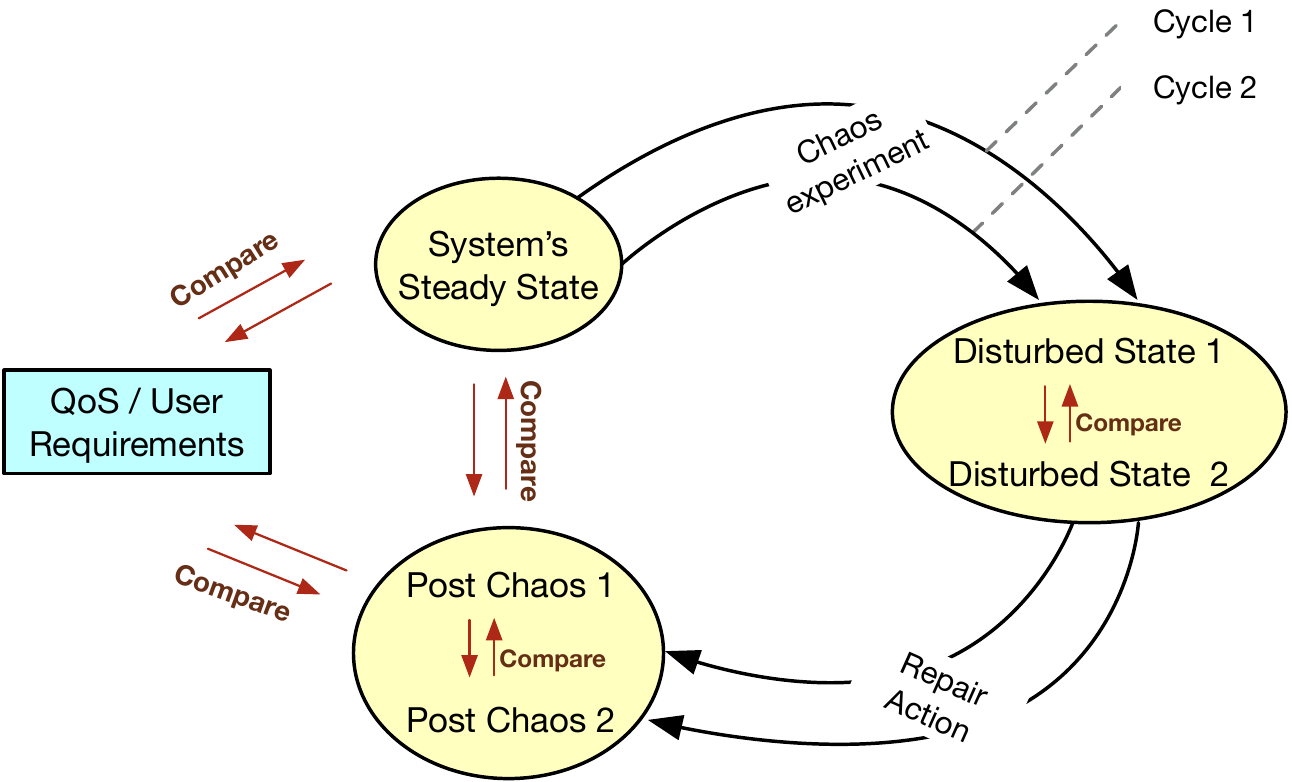}
	\caption{\label{fig:chaoscycle}Comparing results over a sequence of chaos cycles.}
	\vspace*{-3ex}
\end{figure}

To \emph{check the steady-state}, we capture the execution of the system and its components through system logs or execution traces. 
We can \emph{evaluate} the system's behavior based on these logs, user requirements, and software quality attributes.
Moreover, when \emph{designing the chaos experiment}, we can decide the type and level of faults to inject into the system and select a chaos tool accordingly.
During \emph{chaos experiment selection}, we randomly select a chaos experiment to run from the \emph{pool of chaos experiments} to trigger certain system failures.
As a result of these induced failures, the self-healing system would activate repair actions or perform adaptations to keep the system satisfy specific requirements. 
The mechanism will observe and log the system state again to evaluate the system against the selected chaos experiments. 
This process can be repeated in a loop to perform a deeper evaluation covering several aspects and quality attributes. 
Moreover, we can improve the quality of successive chaos experiments by using the \emph{evaluation feedback} against each executed chaos experiment. 
Figure~\ref{fig:chaoscycle} shows different types of comparisons for evaluation as part of the sequential invocations of the chaos cycle.
System states are frequently captured during the chaos cycle. 
This allows for comparisons between various system states both prior to, during, and after the injection of faults, 
as well as compliance checking for the requirements at any given state.

\subsection{Research Questions}

\noindent 
We examine the viability of CHESS through an exploratory study that aims to to address the following research questions: 

\begin{researchquestions}
	\item To what extent can we leverage CE for evaluating SHS and SAS under realistic failure scenarios? \label{itm:CEfaults}	
	\item How does the observability of the system-under-evaluation affect CHESS's evaluation capability? \label{itm:observability}
	\item What kind of limitations does the use of CE bring for the evaluation of SHS and SAS? \label{itm:CElimit}
\end{researchquestions}

\noindent
\ref{itm:CEfaults} aims to understand the nature of failures in SHS and SAS in terms of their properties. 
Moreover, it investigates how we can effectively use CE to model failures that mimic realistic situations and maximize domain coverage. 
In complex distributed systems, different faults can lead to system failures~\cite{avizienis2004:basic}.
Finding the relation between faults and the corresponding failures is often challenging. 
Furthermore, a failure often does not occur in isolation but can be correlated to other failures concerning time or space. 
We can categorize failures into independent failures, uniform or persistent failures, and bursty or cascading failures. 
To develop chaos experiments that %
represent a diverse set of realistic failure scenarios for SHS,
we build on the model for failures occurring in a distributed system~\cite{gallet2010:model}
which proposes three components (i.e., group size, inter-arrival time, and resource downtime). 
Group size refers to the number of failures present in a system in each space-correlated failure
whereas inter-arrival time and downtime refer to the time between consecutive failures and the downtime caused by these failures.       

\ref{itm:observability} explores how limited observability in the context of SHS and SAS affects the evaluation 
and what minimum observability levels are required to ensure evaluation for specific quality attributes.
There can be multiple levels to the observability~\cite{simonsson2021:observability} such as system calls or low-level, component level, and application level.
The lack of observability at these levels can limit our ability to evaluate an SHS.
Furthermore, specific components of a SAS and SHS can be unobservable~\cite{casanova2014:diagnosing, scheerer2021:reliability}. 
Nevertheless, they are often connected to other observable components to perform different tasks.
It is challenging to accurately evaluate the compliance of a particular attribute for a task containing these components. 
Moreover, there can be limitations regarding observability for specific hardware components with SHS deployment or limitations concerning the system's architecture.

Finally, \ref{itm:CElimit} investigates the limitations of using chaos engineering for evaluation, 
and to which extent additional measures can assist in overcoming these limitations. 
For example, chaos engineering can seriously disrupt the functioning of a system~\cite{poltronieri2022:chaos}, 
and the \emph{blast radius} of a chaos experiment is used to describe the extent of the damage that the experiment can cause to a running system. 
This question examines to what extent the blast radius can interact with the need for control and predictability during a systematic evaluation.

\section{Study Design}

\noindent 
We examine the viability of CHESS through an exploratory study on a self-healing smart office application.
We considered several factors while designing this study. 
First, we looked at the most commonly used quality attributes (i.e. performance, availability, reliability, and security) 
in SAS and SHS studies~\cite{weyns2012:integrated, mahdavi-hezavehi2017:systematic}, 
and chose to focus our evaluation on availability, reliability, performance, and integrity. 
Since our smart office does not cover all security aspects, we only considered integrity which is one aspect of security~\cite{villegas2011:framework}. 
There are multiple ways in which researchers have defined these attributes. 
In general, \emph{availability} means when all the components in a given system remain responsive. 
A \emph{reliable} SHS will function according to the specified requirements of the user.  
We can measure a system's \emph{performance} against the time it takes to process a specific event 
whereas \emph{integrity} can be defined as the absence of improper (or unauthorized) system alterations~\cite{avizienis2004:basic}.  
Second, we investigated the models that define and explain failures in the self-healing problem space. 
Several elements exist including failure model, system response, completeness, and design context~\cite{koopman2003:elements}, each corresponding to several factors. 
For instance, the failure model consists of failure duration, failure manifestation, failure source, granularity, and failure profile expectation. 
Ensuring that an evaluation covers all aspects of these elements is often challenging. 
Therefore, we focus instead on perturbations through chaos experiments utilizing the failure model that can lead the system to face diverse scenarios. 
The system can then be observed and evaluated by checking system logs captured throughout the experiments. 
Lastly, we looked at an appropriate environment to design our experiments
that enables varying conditions and the ability to perturb the system and control different levels of observability. 
Therefore, we propose to use microservices-based architecture for our experiment design. 
The following section will discuss the motivation behind choosing microservices and formulate failure scenarios 
that can capture diverse aspects of an SHS's execution, considering the quality attributes under evaluation. 

\head{Distributed systems based on Microservices}
\noindent
Due to the exploitation of benefits like faster delivery, improved scalability, and greater autonomy, 
microservices have become a standard way of implementing modules in SHS~\cite{jamshidi2018:microservices}. 
Although distributed systems with microservices can suffer from network, hardware, or application-level issues and are vulnerable to various factors, 
they are ideal for implementing SHS because microservices architectures offer a variety of strategies for dealing with various issues. 
Moreover, their modular design makes it relatively less complicated to target specific components and keep track of independent units. 
Some existing studies have explored the idea of fault injection and chaos engineering in microservices. 
For instance, FILIBUSTER proposed service level fault injection testing~\cite{meiklejohn2021:servicelevel} 
that systematically identifies resilience issues early in the development of microservice applications. 
ChaosOrca evaluates the resilience of containerized applications through injecting system call errors~\cite{simonsson2021:observability}. 
Frank et al. designed a case study based on microservices architecture for the problem of resilience requirement elicitation 
and used ChaosToolkit-based chaos experiments to assess and improve their architecture~\cite{frank2021:scenario}.  
ChaosTwin is a management framework leveraging chaos engineering to digital twins for improving configurations 
of a cloud-based video streaming service use case~\cite{poltronieri2022:chaos, poltronieri2021:chaostwin}.
Our study considers a distributed IoT system consisting of sensors, actuators, 
and multiple microservices such as rule inference services, control, and system services.

\head{Failure scenarios}
\noindent
Modeling a range of failure scenarios that can cover diverse aspects of the system's behavior and functioning 
aids in performing a systematic evaluation of the system with respect to the quality of services.
Our study leverages a failure model based on previous studies to ensure realistic failures~\cite{koopman2003:elements, gallet2010:model}.  
In addition, we aim to devise categories of failure scenarios that can target a combination of specific quality attributes. 
Our choice of failure scenarios is based on common faults in IoT systems~\cite{norris2022:iotrepair}, as well as the quality attributes considered. 
We consider the following four categories of failure scenarios for performing our evaluation: 

\begin{enumerate}[label=\textit{\textbf{FS-\arabic*:}}, ref=FS-\arabic*, left=\parindent]\itshape
	\item a running service is down abruptly.
	\item a deployed sensor sends erroneous readings.
	\item a deployed sensor is down unexpectedly.
	\item a running service is delayed. 
\end{enumerate}
Table \ref{tab:table-1} shows the mapping of designed categories of failure scenarios to targetting system attributes of availability, reliability, integrity, and performance. 
These failure scenarios affect the quality attributes of specific components in the SHS by exposing the system to different faults and testing to which extent it can provide resilience against these failures. 
Each category covers a range of failure scenarios. 
FS-1 concerns the unavailability of the services, which can cover scenarios ranging from service crashes and service updates to communication disruption between services and timeouts. 
FS-2 can include data corruption, data loss during communication, and a hardware fault causing the component to send incorrect data. 
In FS-3, we can consider the failure of a hardware component (e.g. due to a power outage) resulting in disconnection, or an unauthorized alteration of a hardware component. 
FS-4 may include delays between the services caused by overload at communication channels or delays due to resource exhaustion or limited capacity of a constraint device. 

\begin{table}[t]
	\caption{\label{tab:table-1} Mapping failure scenarios onto system quality attributes}
\centering
	\begin{tabular}{ccccc}
		\toprule
		& Availability & Reliability &  Integrity   &  Performance \\
		\midrule
		 FS-1 & \checkmark &    \checkmark    &      -     & \checkmark \\ 
		 FS-2 &      -     & \checkmark & \checkmark &      -     \\
		 FS-3 & \checkmark &     \checkmark    & \checkmark &      -     \\
		 FS-4 &     \checkmark   &      -     &      -     & \checkmark \\
		 \bottomrule
	\end{tabular} 
	\vspace*{-2ex}
\end{table}

\section{Smart Office Exploratory Study Using CHESS}

\begin{figure}[t]
	\vspace*{1ex}
	\centering
	\includegraphics[width=0.95\columnwidth]{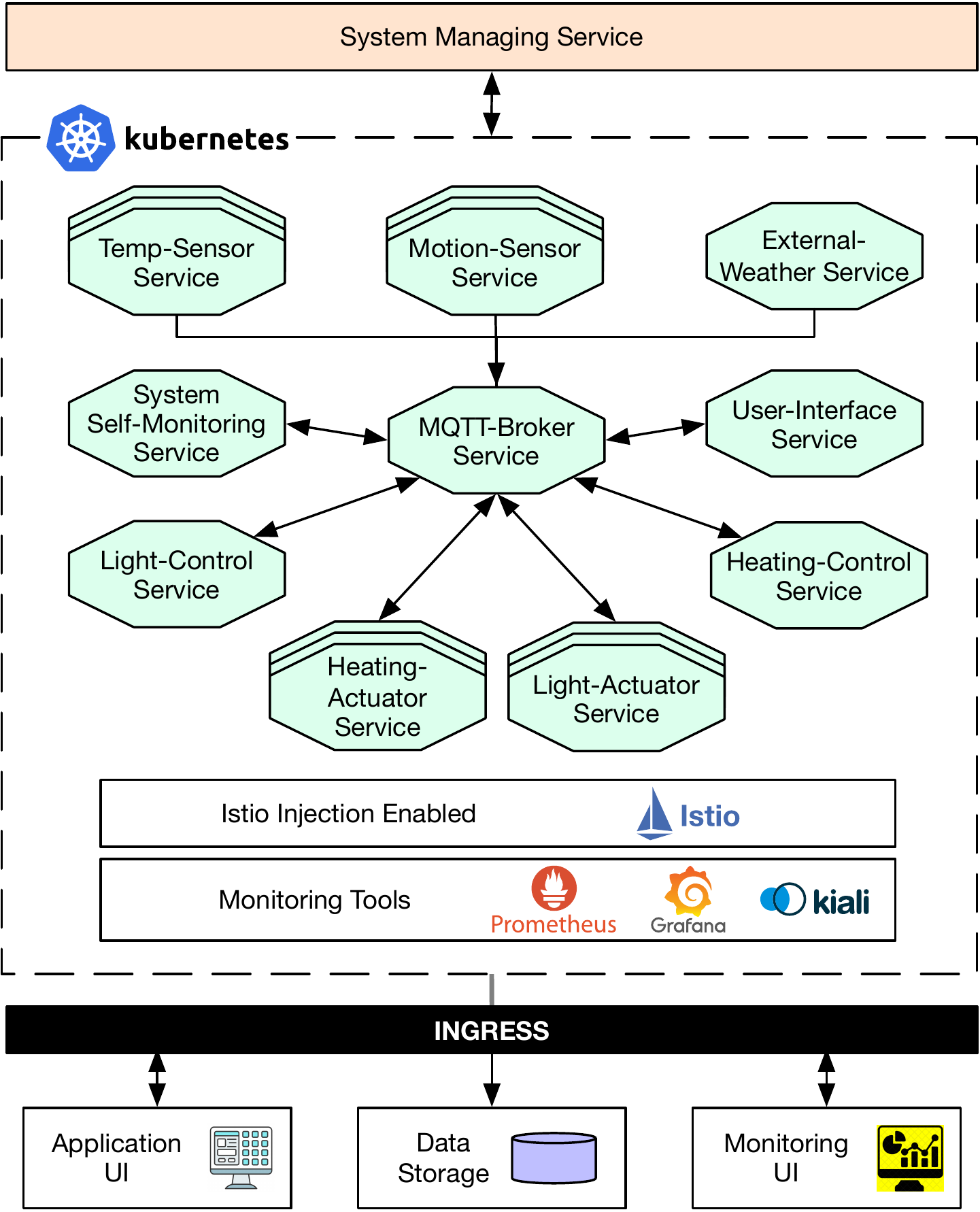}
	\caption{\label{fig:figure-smartoffice}System diagram for deployment of smart office exploratory study}
	\vspace*{-2ex}
\end{figure}

\noindent
In this section, we describe our exploratory study using the framework of CHESS, discuss our approach to cover each failure scenario, and evaluate the implemented system with two rounds of evaluation, 
each addressing different aspects of evaluation. 
We consider a self-healing smart office for deployment, experimentation, and monitoring. 
We consider a simple scenario with two types of sensors, i.e., temperature and motion sensors, two types of actuators, i.e., light and heating actuators, and one external service for the weather. 
The smart office scenario follows a set of user requirements such as indoor temperature range [min, max], illumination levels based on time of the day and weather conditions, 
and timeout for switching off lights when the user is not present. 
The deployed smart office services are sensor, external weather, control, actuator, and user interface. 
We have two additional services named system monitoring and system managing, where the former covers fault detection and the latter covers fault diagnosis and recovery. 
All these services interact with each other via the MQTT broker. 
The sensor services publish periodic sensing data from the sensors to the MQTT broker, and the external weather service publishes the periodic weather data to the broker. 
The control services take the sensing and weather data as input and decide the control values for light and heating based on current conditions and user-defined preferences. 
The user interface shows the incoming sensing values; events generated, weather conditions, current actuator controls, and sensors battery levels. 
The system monitoring service keeps track of the running states of all services, and the system managing service performs edit actions on the running services. 
The edit actions include deploying new services, deleting services, and updating service configuration files. 
Figure \ref{fig:figure-smartoffice}  shows the overall architecture of the deployment and interactions of the service.

We designed sets of \emph{chaos tests} for each failure scenario following a failure model~\cite{gallet2010:model} 
and chose ChaosToolkit as the chaos engine for performing chaos experiments on our system. 
ChaosToolkit works with containers, Kubernetes, bare metal, and most cloud providers. 
It comes with extensive documentation and an example experiments suite. 
In our scenario, we use ChaosToolkit-Kubernetes, one of the several extensions of ChaosToolkit. 
We deploy the smart office services and the self-monitoring service within a Kubernetes cluster, 
allocating 4 CPUs and 6000MB memory.\footnote{~A replication package containing a VM for CHESS 
and the Smart Office exploratory study is available via \url{https://doi.org/10.5281/zenodo.6817764}.}
The number of CPUs and memory needed may differ for another managed system, depending on the number/type of services to be run.

\subsection{Failure Scenario 1: Service Down}

\noindent
\noindent
In this failure scenario, we design a set of \emph{chaos tests} that makes a running service unavailable by terminating its running pods.
\emph{Steady state hypothesis:} all the running microservices should be in healthy condition and responsive.
A service can be made unavailable for a particular time by deleting its running pods back-to-back. 
The SHS deals with this failure by adding auto-scaling to the service, which allows the service to keep multiple replicas based on available CPU resources. 
The chaos test includes:  (i) deleting back-to-back pods for each service one by one, (ii) deleting pods for a combination of \emph{k} services, 
and (iii) repeating the first two sets of experiments with varying time intervals between them.

\subsection{Failure Scenario 2: Sensor Fault}

\noindent
In this failure scenario, we design a set of \emph{chaos tests} that makes a deployed sensor service faulty by injecting false readings into it.
\emph{Steady state hypothesis:} the control service should not be receiving erroneous sensing data through data validation.
The SHS deals with this failure by adding data validation to the responsible services, and checking if the data is realistic based on given ranges.
It allows the SHS to timely identify, unsubscribe, delete the faulty service, and replace it with the deployment of a healthy service.  
We consider the injection of realistic and unrealistic erroneous readings.
Note that realistic false readings are harder to detect due to their similarity with the correct readings.
Our set of chaos tests includes:  
(i) injecting the same realistic false readings to all the sensor services of one type, 
(ii) injecting the same unrealistic false readings to all the sensor services of one type,
(iii) injecting mixed realistic and unrealistic readings to all the sensor services of one type,
(iv) repeating the first three experiments to multiple types of sensor services.  

\subsection{Failure Scenario 3: Sensor Down}

\noindent
In this set of experiments, we design a set of \emph{chaos tests} that makes a sensor down. 
\emph{Steady state hypothesis:} the sensor service must not be in an idle state. 
We can make a sensor unavailable by making its battery resource drain. 
The SHS deals with this failure by checking the sensor battery and connection status via the responsible services, 
and raising an alarm for the replacement of the sensor when battery drainage is detected.
If a backup sensor is already available, it enables it to respond promptly by connecting the sensing service to the backup sensor. 
Our set of chaos tests includes:  
(i) draining the battery of sensors one by one, 
(ii) draining the battery of a combination of \emph{k} sensors, 
and (iii) repeating the first two sets of experiments with varying time intervals between them.

\subsection{Failure Scenario 4: Service Delayed}

\noindent
In this failure scenario, we design a set of \emph{chaos tests} that reflects a delay in the communication of running services. 
\emph{Steady state hypothesis:} a relevant service must respond to the sent commands within a set delay threshold.
We can slow the service communication by injecting a periodic delay into it, e.g., injecting a delay of 20 seconds in the communication after every two minutes.
The SHS deals with this failure by keeping track of the average response times of the running services. 
It aids in identifying service delays at early stages, terminating services that exceed the threshold, and replacing them by deploying healthy service instances. 
The chaos test includes:  
(i) injecting delay in the services one by one, 
(ii) injecting delay in a combination of \emph{k} services, 
and (iii) repeating the first two sets of experiments with varying time intervals.

\subsection{Evaluation}

\noindent
In order to evaluate our system, we perform two rounds of chaos experiments 
to examine changes in system behavior and their impact on quality attributes. 
The first evaluation round examines \emph{self-healing behavior}: 
we perform failure injections into the running services that allow us to both analyze the impact of the failures through the \emph{blast radius}, 
as well as analyze the effects on the system's quality attributes referenced in Table \ref{tab:table-1}. 
The second evaluation round examines \emph{self-adaptive behavior}: we test our system under increasing loads to evaluate how these affect the performance of the system.
\begin{enumerate}[label=\textit{Round \arabic*:}, ref=Ex-\arabic*, left=0pt]\itshape
	\item (self-healing) single-service/multi-service failure injection to examine blast radius.
	\item (self-adaptive) step-wise increase in service requests to examine performance under varying loads.
\end{enumerate}

\begin{figure}[t]
	\vspace*{1ex}
	\centering
	\includegraphics[width=0.95\columnwidth]{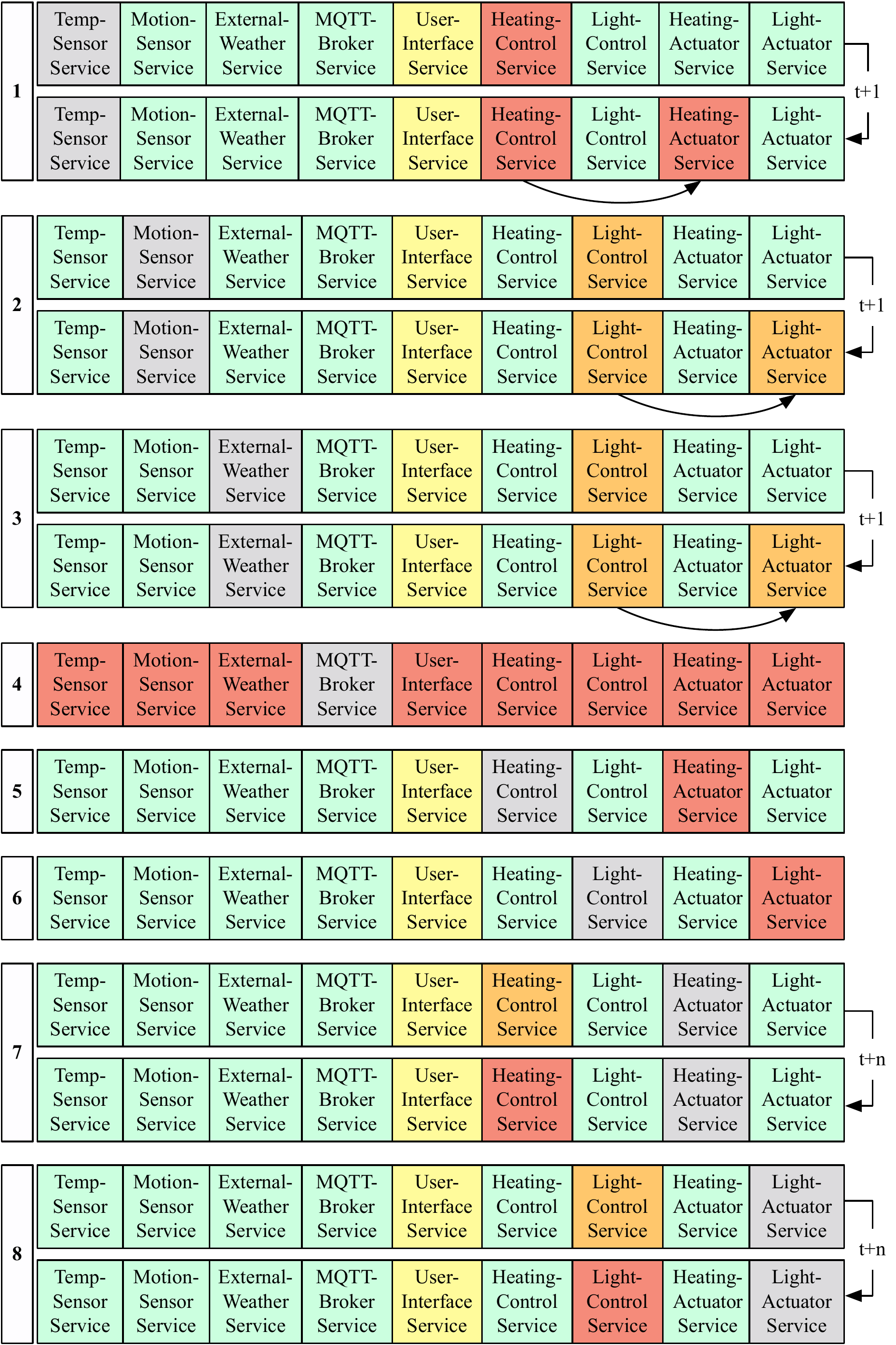}
	\caption{\label{fig:figure-Blast}Blast radius of chaos experiments on vulnerable services}
	\vspace*{-2.2ex}
\end{figure}

\head{Self-healing Evaluation (Round 1)}
We inject failures into eight services and show how a deviation in each service's running state impacts other services.
Figure~\ref{fig:figure-Blast} shows the blast radius of the system for running chaos experiments on the vulnerable deployed services. 
 The grey color represents the primary service of the failure injection in a chaos experiment. 
The yellow color represents low-level impact (functionality slightly affected), 
the orange color represents medium-level impact (functionality partially affected), 
and the red color represents high-level impact (service not functional).

We can map failure scenario 1 (FS-1) to each of the services, and the resulting impact is shown in Figure~\ref{fig:figure-Blast}, rows 1 to 8. 
Failure scenario 2 (FS-2) and failure scenario 3 (FS-3) correspond to the temperature sensor service failure (row 1) and motion sensor service failure (row 2). 
Similarly, failure scenario 4 (FS-4) also maps to each of these services, i.e., from sensor service delay to control and actuator service delay. 

Figure~\ref{fig:figure-Blast}, row 1, shows the blast radius for when the temperature service deviates from its running state. 
When the temperature sensor service faces a failure, at first, 
it puts a low-level impact on the user interface service and a high-level impact on the heating control service. 
User interface service has a low-level impact because all other services are still running, and the service data is being streamed. 
However, the user cannot visualize the temperature and resulting temperature control data. 
A high-level impact on heating control service because it highly depends on current temperature values to devise decisions for heating control. 
At the next timestamp, there is an impact on the heating actuator service as it depends on the control commands from the heating control service. 

Figure~\ref{fig:figure-Blast}, row 2, show the blast radius for when the motion sensor service deviates from its running state. 
In this case, the light control service only has a medium-level impact because our system derives the light control from motion event information, outdoor weather conditions, and the current time of the day. 
When one of the inputs is unavailable, the light control is based on the other two available inputs. 
Hence, the light control service is still functional but not to its full potential.
In the next timestamp, the light actuator service is impacted due to the impact on the light control service. 
The blast radius of the external-weather service (Figure~\ref{fig:figure-Blast}, row 3) is similar to that of the motion sensor service. 

Figure~\ref{fig:figure-Blast}, row 4, shows that all running services are down as the result of a failure of the MQTT-broker service. 
In Figure~\ref{fig:figure-Blast}, row 5 and row 6, the blast radius for a failure in heating control 
and light control reflects a high-level impact on the heating actuator service and light actuator service, respectively. 
Finally, Figure~\ref{fig:figure-Blast},  row 7 and row 8, show that the blast radius for a failure in the heating actuator service 
and light actuator service at first reflects a medium-level impact on heating control service and light control service, respectively. 
The control services stay functional if the actuator services face a failure for a short period. 
However, if the actuator services remain down for a significant amount of time, it eventually puts a high-level impact on control services. 
We observe that the blast radius for service delay failures varies based on the average delay time.

Further, we observe whether the failure of one service can make other services unavailable, less responsive, delayed, faulty, or idle. 
None of the failure scenarios deviates from any other service's state to unavailable. 
All four failure scenarios make other services less responsive and delayed due to either no input data from a service, incorrect input data, or delayed response from the service. 
Only FS-2 can make other services faulty, as when incorrect sensing data is received, it can lead to the generation of faulty control commands. 
All four failure scenarios can deviate other services' states to idle for a certain amount of time, e.g., when there is no input data, delayed input data, or incorrect input data. 

\begin{figure}[t]
	\centering
	\includegraphics[width=0.95\columnwidth,trim=0 39pt 0 10pt, clip]{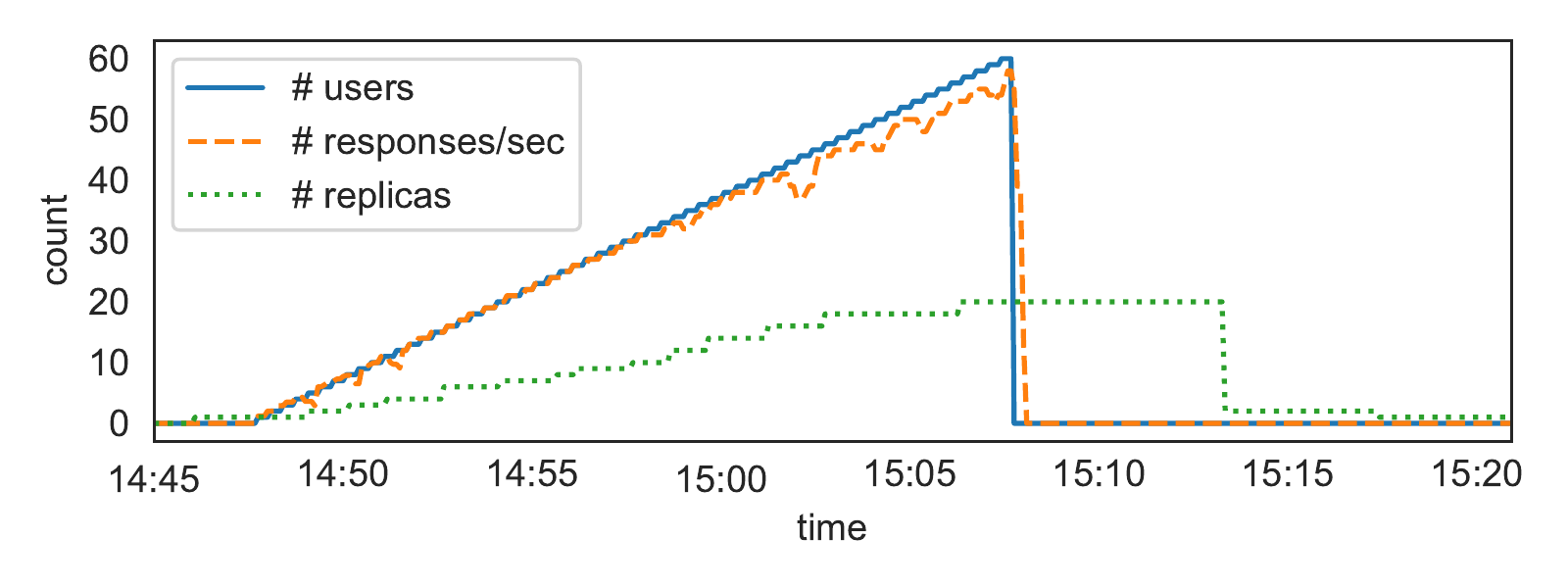}
	\hspace*{-2pt}\includegraphics[width=0.958\columnwidth,trim=0 14pt 0 0pt, clip]{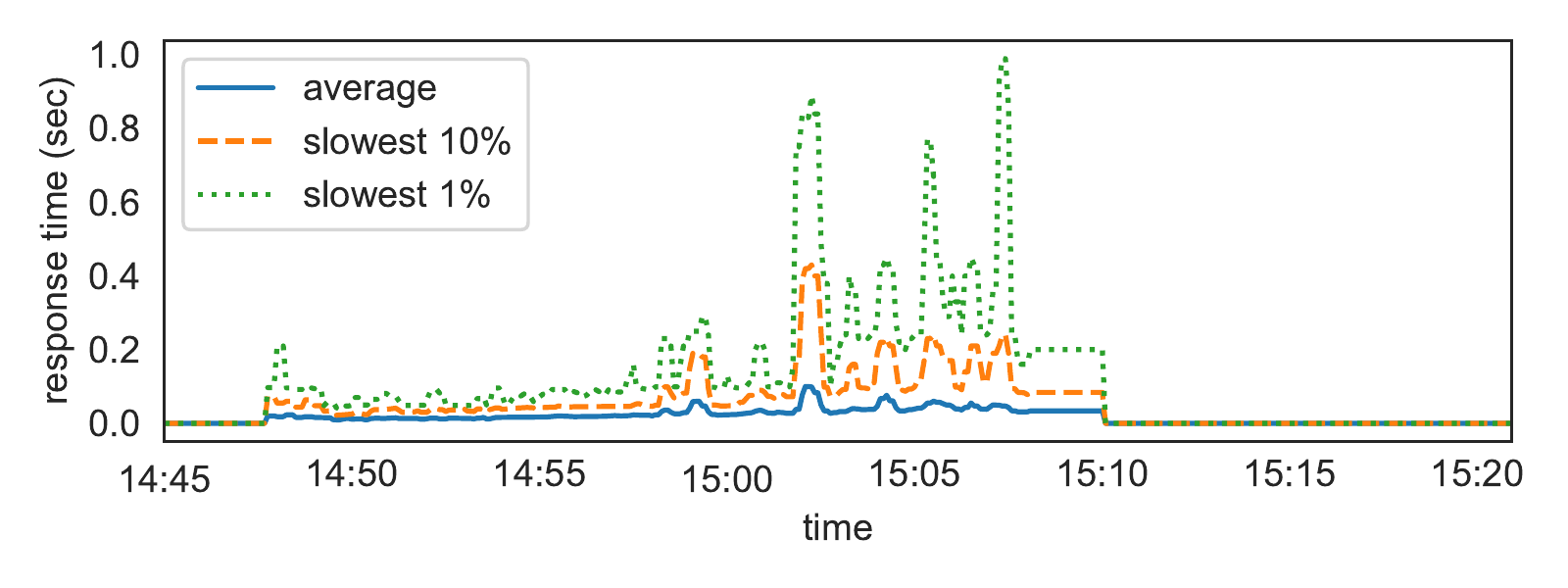}
	\begin{picture}(200,0)(-2,0)%
		\put(45,92){\tiny \textsf{duration of chaos experiment}}
		\put(43,93){\vector(-1,0){24}}
		\put(110.5,93){\vector(1,0){24}}
		\linethickness{.1pt}
		\multiput(19,100)(0,-2){7}{\line(0,-1){1}}
		\multiput(19,59)(0,-2){16}{\line(0,-1){1}}
		\multiput(134.5,100)(0,-2){37}{\line(0,-1){1}}
	\end{picture}		
	\vspace*{-3ex}
	\caption{\label{fig:performance}Performance evaluation under varying system load}
	\vspace*{-2ex}
\end{figure}
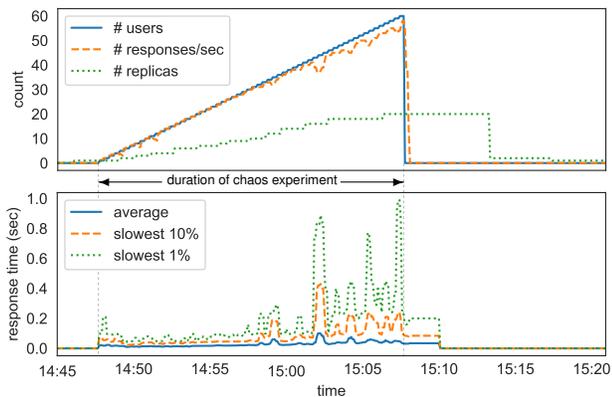

\head{Self-Adaptive Evaluation (Round 2)}
We analyze the performance of a system under load using an experiment that gradually increases the service request rates. 
We run the experiment of gradually increasing the load for twenty minutes. 
The experiment starts with one user and adds a new user every 20 seconds. 
Each user attempts to access the light control service once every second. 

The top graph of Figure~\ref{fig:performance} shows the load and capacity of the system, represented by the number of users, the number of service responses per second, 
and the number of deployed service replicas.
The bottom graph shows the system response times corresponding to the load, represented by the average response time, 
response time of the slowest ten percent requests, and the response times of the slowest one percent requests. 
The X-axis shows the timeline of the experiment.

The top graph shows that as the number of users starts rising, there is a corresponding rise in the number of responses per second.  
The service starts with one replica running, but the number of replicas gradually increases with the increase in incoming requests' load. 
At the experiment's peak, the number of replicas reaches twenty, which was the configured maximum. 
After termination of the user load, the number of available replicas remains twenty for the following seven minutes, 
then reducing to two replicas for the following four minutes before it gets down to one replica again. 
The system delays reducing the number of replicas as a safe strategy for maintaining buffer time so it can execute any pending requests and an unexpected rise in the load. 
At the end of the experiment, we see that the number of responses does not drop suddenly, reflecting a few seconds delay in the service response under increased load.

The bottom graph shows that the response time starts with a relatively gradual increase during the first half of the experiment, 
while there is still less load on the service. 
High variations are observed in the response times during the second half of the experiment, as the loads keep increasing every twenty seconds,
with more users trying to access the service each second. 
Once the active users are terminated, the response time drops from 1 second to 0.2 seconds. 
It remains 0.2 seconds for about two minutes, which reflects the handling of any pending tasks before dropping down to zero.

\section{Lessons learned}

\noindent
This section summarizes the lessons learned from the experiments that evaluate a self-healing smart office using CHESS.

\head{\ref{itm:CEfaults}} Chaos engineering can be used to inject multiple type of faults into the containerized applications, that can mainly be divided into two main catergories of  \emph{infrastructure level faults}  and \emph{functional level faults}.
Chaos engines (and their platform-specific extensions) contain predefined chaos directives that can be used to inject \emph{infrastructure level faults} by modifying details about the platform configuration, such as adding, deleting, and scaling deployments and services, rerouting communication, and killing endpoints. 
On the other hand, \emph{functional level faults} require additional knowledge of the system's backend logic and are therefore not directly supported by pre-defined chaos directives. The chaos engines provide a solution of custom faults injection for such cases.
This challenge can be overcome by 
(i) extending the system with custom functions that trigger such functional level faults, and 
(ii) using specific functionality of the chaos engine to call these custom functions as part of a chaos experiment.
Table~\ref{tab:table-3} presents the fault injection levels for the various failure scenarios.
Depending on the failure, a system may require infrastructure level, functional level, or both.
FS-3 (Sensor Down) is a special case as it can be handled by adding a custom function that changes a service's behavior, or by adding a second (faulty) service that is replacing the correct service as the result of an infrastructure modification.

\begin{table}[t]
	\caption{\label{tab:table-3}Levels of fault injection needed for the failure scenarios}
	\centering
	\begin{tabular}{lcc}
		\toprule
		& Infrastructure  &  Functional \\ 
		& Level  &  Level \\ 
		\midrule
		FS-1 : Service Down  & \checkmark &      -     \\ 
		FS-2 : Sensor Fault & \checkmark & \checkmark \\
		FS-3 : Sensor Down & \checkmark & (\checkmark)  \\
		FS-4 : Service Delayed & \checkmark &      -     \\
		\bottomrule
	\end{tabular} 
	\vspace*{-2ex}
\end{table}

\head{\ref{itm:observability}} The levels of \emph{observability}, for a system, heavily depend on the type of implemented services e.g. observability level for containerized applications is mostly in black-box manner, i.e., the internals of a deployed service can not be monitored.
The \emph{monitoring tools} (such as prometheus, kiali and grafana) enable observation of metrics regarding running deployments and services including traffic inflow, traffic outflow, services' health, and infrastructural level parameters such as deployments' impact on system resources (e.g., RAM, CPU) and network load. 
During the chaos experiments, the chaos engine also collects \emph{chaos logs} that collect information regarding exploited services and changes in the system's running state along with the metrics like number of pods available, killed, or terminated.
However, many system details at the functional level cannot be observed using monitoring tools or chaos logs, 
for example, determining if a service is receiving erroneous data or if a service is stuck in a certain state due to a functional error. 
We address this shortcoming in CHESS using \emph{system self-monitoring} which ensures that in-depth observability is available.
This level of monitoring also serves to increase the confidence in the evaluation results obtained. 
Table~\ref{tab:monitoring} shows a comparison of the various monitoring options for CHESS experiments and the extent to which certain failure scenarios need the observability provided by these monitoring options. 

\head{\ref{itm:CElimit}} One of the main challenges while using CE for evaluating SHS is to control the \emph{cascading effects of chaos experiments}.
For a systematic and thorough evaluation, it is vital to avoid superfluous failures.
However, limiting the blast radius to the set of relevant components requires an in-depth understanding of system functionalities and dependencies.
The following measures can help to improve a system's capacity to limit the blast radius:
\begin{enumerate}
	\item \emph{Context-awareness} can support effective decision-making and enable more effective evaluations of SHS. 
	\item \emph{Priority-awareness} of involved tasks can help with addressing the most critical failures first, 
		resulting in a more optimal flow of resolving failures. This is especially important in case of multiple failures and can help prevent cascading failures,
		thereby minimizing the blast radius. 
	\item \emph{Conditional monitoring} allows one to dig deeper into error-prone areas while keeping the monitoring overhead under control, 
		aiding in fault identification and diagnosis.
\end{enumerate}

\head{Summary}
Our overall conclusion is that the proposed CHESS approach is viable, and that it enables systematic evaluation of a self-adaptive or self-healing system by exposing the system to a series of systematic perturbations at both the functional and infrastructural level and analyzing its capacity to maintain, or return to, its steady-state. 

\begin{table}[t]
	\caption{\label{tab:monitoring} A comparison of monitoring options for CHESS experiments}
	\centering
	\begin{tabular}{lccc}
		\toprule
		&  Monitoring &  Chaos  &  System \\
		&   Tools     &   Logs  &  Self-Monitoring \\
		\midrule
		FS-1 : Service Down       & \checkmark & \checkmark &      -      \\ 
		FS-2 : Sensor Fault     &      -     &     -      &  \checkmark \\
		FS-3 : Sensor Down & \checkmark & \checkmark &  \checkmark \\
		FS-4 : Service Delayed   & \checkmark & \checkmark &      -      \\
		\bottomrule
	\end{tabular} 
	\vspace*{-2ex}
\end{table}

\section{Concluding Remarks}

\head{Contributions}
This paper presents \emph{CHESS}, an approach for the systematic evaluation of self-adaptive and self-healing systems that builds on chaos engineering principles. 
The approach was informed by our literature survey of the state-of-the-art in evaluating self-adaptive and self-healing systems, which distinguishes the main evaluation approaches used in the self-adaptive and self-healing literature and highlights the main quality attributes analyzed in those evaluations. 
CHESS systematically perturbates the system-under-evaluation and records how the system responds to those perturbations.
We present the experimental design for evaluating distributed SHS based on microservices and discuss common failure scenarios and their mapping to quality attributes. 
We investigate the viability of the proposed approach through an exploratory study that evaluates the resilience of a self-healing smart office application.
We discuss the \emph{lessons learned} while conducting the study, which includes challenges w.r.t. chaos experiments at the functional level (RQ1), the need for observability at various levels of abstraction (RQ2), and limiting the cascading effects of chaos experiments (RQ3). 
We conclude that with the help of CHESS, we can analyze the system's ability to maintain a steady-state under adversarial perturbations at both functional and infrastructural levels.
Consequently, CHESS enables us to effectively and systematically evaluate the proper behavior of self-adaptive and self-healing systems. 

\head{Directions for Future Work}
We are extending our work with CHESS for data synthesis, which can provide training data for self-healing systems with AI components.  
Other directions for future work include evaluating the application of CHESS to complex architectures such as multi-level interdependent microservices and selecting chaos experiments with the combination of multiple failure scenarios.
It could also be interesting to investigate techniques that enable automated region selection for chaos experiments based on the health and performance status of services.
Further options to extend this work include adding contextual information about the system-under-evaluation, for example using ontologies or knowledge graphs, that can be exploited in chaos experiments for smart targeting of fault injection.
Finally, for large and complex systems, techniques similar to test-case selection may be needed for selecting chaos experiments. 
More research is needed to analyze how to do this without affecting evaluation quality.

\medskip
\head{Acknowledgements}
This work is supported by the Research Council of Norway through the cureIT grant (\#300461).
Sehrish Malik was in part supported through an ERCIM Fellowship.

\medskip

{\balance
\sloppy
\printbibliography
}

\end{document}